\documentclass[manuscript]{acmart}
\usepackage{graphicx}
\usepackage{amsmath, amstext}
\usepackage{caption}
\usepackage{multirow}
\usepackage[thinlines]{easytable}
\usepackage{wrapfig}
\usepackage{subcaption}
\usepackage{bbm}
\usepackage{hyperref}
\usepackage{cleveref}
\usepackage{bm}

\AtBeginDocument{%
  \providecommand\BibTeX{{%
    \normalfont B\kern-0.5em{\scshape i\kern-0.25em b}\kern-0.8em\TeX}}}

\setcopyright{acmcopyright}
\copyrightyear{2022}
\acmYear{2022}
\acmDOI{}

\acmConference[TRAIT at CHI 2022]{Workshop on Trust and Reliance in AI-Human Teams at CHI 2022}{April 30,
  2022}{New Orleans, LA}
\begin{document}

\title[User Trust on Explainable AI-based Medical Diagnosis Support System]{User Trust on an Explainable AI-based Medical Diagnosis Support System}

\author{Yao Rong}
\email{yao.rong@uni-tuebingen.de}
\affiliation{%
  \institution{University of T\"ubingen}
  \streetaddress{Sand 14}
  \city{T\"ubingn}
  \country{Germany}
  \postcode{72074}
}

\author{Nora Castner}
\affiliation{%
  \institution{University of T\"ubingen}
  \streetaddress{Sand 14}
  \city{T\"ubingn}
  \country{Germany}
  \postcode{72074}
  }
\email{nora.castner@uni-tuebingen.de}

\author{Efe Bozkir}
\affiliation{%
  \institution{University of T\"ubingen}
  \streetaddress{Sand 14}
  \city{T\"ubingn}
  \country{Germany}
  \postcode{72074}
 }
\email{efe.bozkir@uni-tuebingen.de}

\author{Enkelejda Kasneci}
\affiliation{%
  \institution{University of T\"ubingen}
  \streetaddress{Sand 14}
  \city{T\"ubingn}
  \country{Germany}
  \postcode{72074}
  }
\email{enkelejda.kasneci@uni-tuebingen.de}





\renewcommand{\shortauthors}{Rong, et al.}

\begin{abstract}
Recent research has supported that system explainability improves user trust and willingness to use medical AI for diagnostic support. In this paper, we use chest disease diagnosis based on X-Ray images as a case study to investigate user trust and reliance. Building off explainability, we propose a support system where users (radiologists) can view causal explanations for final decisions. After observing these causal explanations, users provided their opinions of the model predictions and could correct explanations if they did not agree. We measured user trust as the agreement between the model's and the radiologist's diagnosis as well as the radiologists' feedback on the model explanations. Additionally, they reported their trust in the system. We tested our model on the CXR-Eye dataset and it achieved an overall accuracy of 74.1\%. However, the experts in our user study agreed with the model for only 46.4\% of the cases, indicating the necessity of improving the trust. The self-reported trust score was 3.2 on a scale of 1.0 to 5.0, showing that the users tended to trust the model but the trust still needs to be enhanced. 
\end{abstract}




\begin{teaserfigure}
  \centering
  \includegraphics[width=\textwidth]{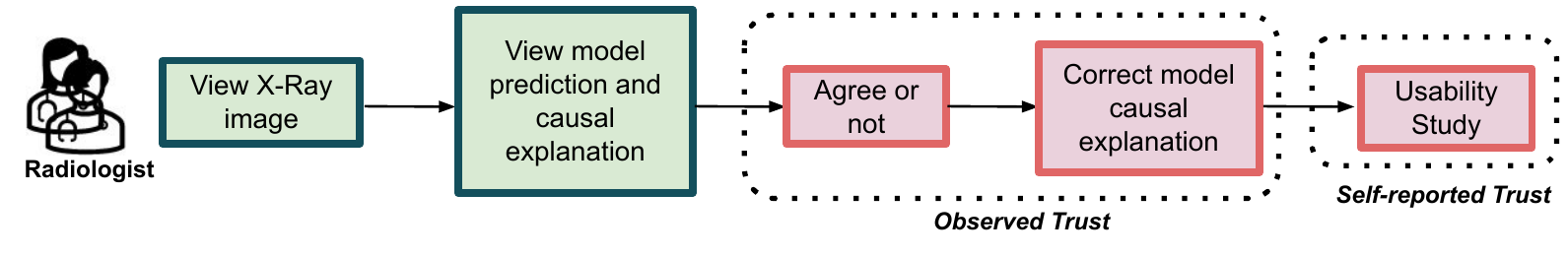}
  \caption{\textbf{Workflow of our study:} We designed and trained an explainable AI model for Chest X-ray images. We measured the user trust and reliance based on their agreement with model explanations and final decisions (\textit{observed trust}). Moreover, expert opinion on usability was asked to report their trust using the model and AI in general (\textit{self-reported trust}).}
  \label{fig:teaser}
\end{teaserfigure}
\maketitle

\section{Introduction}
\label{sec:intro}

There is an ever-growing interest in employing AI-based Medical Diagnosis Support Systems (AIMDSS). However, there is often hesitation when it comes to actual integration into clinical environments~\cite{amershi2019guidelines,budd2021survey}. Medical professionals cannot afford the extra burden of a system that may react unpredictably and, more important, they may not follow its logic. When AI interaction is uncomfortable for the user, trust is diminished regarding the system's performance and practicality~\cite{amershi2019guidelines}. Recently, more effort has been directed towards improving the user trust in AIMDSS 
\cite{bussone2015role,larasati2020effect,xie2020chexplain,schoonderwoerd2021human}.
\textit{Trust} is defined in \cite{lee2004trust} as \textit{``the attitude that an agent (system) will help an individual (user) achieve their goal in a situation that creates uncertainty and vulnerability for the individual"}. Other definitions of trust are more subjective: For instance, trust can be the user's faith that the model will perform well, or when the user feels comfortable working with a well-understood model~\cite{lipton2018mythos,mohseni2021multidisciplinary}. As eXplainable AI (XAI) techniques emerged, previous works \cite{bussone2015role,kim2015interactive,larasati2020effect,mohseni2021multidisciplinary} define trust as the comfortable and confident feeling fostered when using a system. They suggested using explainable AI models to earn the trust of users. In this paper, we study user trust based on their responses to a model that presents explanations of its final diagnosis. Users could also alter these explanations, which would update the model's diagnosis. 

Trust metrics used in mush of the existing research can be divided into two groups: \textit{self-reported} and \textit{observed} trust \cite{papenmeier2019model}.
Self-reported trust is measured in the form of questionnaires or interviews with scores from on Likert scale, i.e., users rate how confident they feel when using the model~\cite{bussone2015role,larasati2020effect,holliday2016user,sollner2012use,pu2006trust,papenmeier2019model,radensky2021exploring,buccinca2021trust,yin2019understanding,ribeiro2016should,ribeiro2018anchors}. 
Since trust includes different factors and they make \textit{trust} more concrete and comprehensive -- such as perceived reliability, perceived understandability, and faith \cite{madsen2000measuring} -- it is measured based on multiple questions that address these different facets \cite{larasati2020effect,papenmeier2019model}. Self-reported trust is crucial in understanding experts' feelings when interacting with the system. 
Observed trust is measured by observing human behavior when they use the model, which offers understanding of how they interact with the system~\cite{radensky2021exploring,buccinca2021trust,yin2019understanding,ribeiro2016should,ribeiro2018anchors}. For instance, \cite{yin2019understanding,zhang2020effect} used the agreement fraction (the fraction where the subject’s final prediction agrees with the model’s prediction) and switch fraction (the fraction where subject aligned their predictions with the model after seeing the model's prediction) to indicate human trust in a simple classification task. Observed trust metrics are necessary since the self-reported trust may not be reliable \cite{zhang2020effect} and even contradictory~\cite{papenmeier2019model}. 
Therefore, for more comprehensive insight to user trust, this paper is measuring trust as a combination of both self-reported and observed metrics. 

The level of trust determines how \emph{reliant} the user becomes on the system's output~\cite{lee2004trust}. \cite{ribeiro2016should,ribeiro2018anchors,radensky2021exploring} also propose that explainable models help determine the appropriate level of trust. 
Too high of trust level can lead to \emph{over-reliance}, where incorrect model output is overlooked due to too much faith in the system. Too low if trust level can lead to \emph{self-reliance}, where model suggestions -- even if correct -- are ignored, which defeats the purpose of even having a support system. 
Bussone et al.~\cite{bussone2015role} found increased explainability from a diagnosis decision system improved the user trust, but also led to \textit{over-reliance}, while less explainability harmed trust, which led to \textit{self-reliance}. 
In this paper, we gave users the option to change the model explanations and use the users' deviation from the model as an observed indicator for trust. By measuring these applied changes, we can determine the level of trust and how it aligns to over- and self-reliance.
The workflow of our project is illustrated in figure \Cref{fig:teaser}. We let radiologists first view the image and then the model prediction as well as its causal explanation.
Causal explanations refer to predictions of anomalies (e.g., lung opacity or lung lesion) that leads to the final diagnosis (e.g. congestive heart failure). After viewing, medical experts choose whether they agree with the model diagnosis or not, then they can change the causal explanations. 
Based on experts' opinions collected in the usability study, we can gain insight into the user - trust as well as their observed interaction with the model. 
The goal is to improve user trust and better human-AI collaboration in the context of medical decision-making by involving the expert. 
It is pivotal to incorporate professional feedback into the development of these systems, because ultimately these are the people who will be employing them. We want experts to \emph{want} to use these systems. 

Our contributions are as follows: We designed and trained an AI model for diagnosis based on chest X-ray images, which offers causal explanations to support its diagnoses. Then, we conducted a user study among experts to investigate the trust and usability of our model. Based on this user study, we evaluate:
(1) Whether medical professionals trust our model. (2) Whether medical professionals are over-reliant or self-reliant, or have appropriate reliance. (3) How we can better design the model to increase and calibrate the trust of users.




\section{Related Work}
\paragraph{XAI and user trust.} Explainable AI plays an essential role in helping users to understand and appropriately trust AI models \cite{gunning2019darpa}. Recent research have studied how the type or the quality of explanations can affect the user trust on different tasks. For example, \cite{papenmeier2019model} studied a text classification model on detecting whether a Tweet was offensive or not and found that adding explanations more often harmed the user trust, especially when the explanations were low-fidelity. \cite{radensky2021exploring} explored the different effect on user trust when providing global and local explanations in a research-paper recommendation system. It suggested that both explanations rather than either alone could help users better understand the system. Moreover, \cite{zhang2018predicting} aimed to examine the effect of local explanations on the trust calibration in an income prediction task. Unfortunately, they found out that explanations did not have impact in calibrating trust. Therefore, individual user studies on trust are necessary especially in medical domain, where inappropriate trust can result in misdiagnosis \cite{bussone2015role,larasati2020effect}.

\cite{bussone2015role,larasati2020effect} focused on how different explanation styles influenced the user trust in medical applications. Different styles refer to how the explanations are presented, for example, detailed explanations v.s. less detailed explanations in \cite{bussone2015role}, or contrastive explanations in \cite{larasati2020effect} where it explained as "why A and why not B". 
However, both user studies were conducted using simulated (fictional) data and explanations ("Wizard of Oz" approach), and only included self-reported trust measurement. Another drawback of the user study in \cite{larasati2020effect} was that the participants were not from medical domain (non-experts). Compared to both previous works, our work used real-world data from a trained model and conducted the user study with the medical experts, which makes the results more convincing. Our user study also measured the observed trust as the deviation between humans' and model's explanations, as we found this suitable for the complexity of the current medical diagnosis task.
\paragraph{AI models for chest X-ray interpretation.} 
Since chest X-ray images are one of the most frequent medical images in practical use, the accurate, automated analysis of chest X-ray images is becoming increasingly of interest to researchers \cite{johnson2019mimic,wang2017chestx}. There are several very large chest X-ray datasets: MIMIC-CXR \cite{johnson2019mimic}, Chest-Xray8 (NIH) \cite{wang2017chestx}, and CheXpert \cite{irvin2019chexpert}, to name only a few. The images in these datasets are labeled with approximately thirteen diagnoses~\footnote{Each dataset has slightly different labels} that can be used to train multi-label classifiers~\cite{seyyed2020chexclusion,yao2017learning,rajpurkar2018deep,akbarian2020evaluating}. Recently, the eye gaze data of a radiologist when reading the X-ray images was published in CXR-Eye dataset~\cite{karargyris2021creation}. In addition, they collected three-class clinical observations from professionals in the Emergency Department (ED prognosis). Our proposed system for chest X-ray inspection differs from conventional, black-box AI tools because our model provides the causal explanation of its decision. Inspired by \cite{koh2020concept}, we used the thirteen diagnoses (anomalies) to realize an ``explaining" layer for predicting a final prognosis (ED prognosis) given in the dataset CXR-Eye. 
\section{Methodology}
Our proposed work aims to find out whether users trust our explainable model and assesses their level of trust. We introduce the model's design, the causal explanations, and how it was trained in \Cref{sec:explainable model}. Then, the user study design for the trust assessment, data collection, and analyses are detailed in \Cref{sec:study design}.

\subsection{Explainable Model}
\label{sec:explainable model}
The system is formalized as follows: Given a chest X-ray image input $X \in \mathcal{R}^{3 \times {H}\times{W}}$, an encoder $\mathcal{E}(\cdot)$ encodes the input image $X$ into the feature vector $v$. The explainable layer is denoted as $f_e(\cdot)$, the final prediction (ED prognosis) layer as $f_c(\cdot)$. First, the explainable layer gives the probability of each anomaly $\hat a = f_e({v})$, i.e. causal explanations. Then, the final prediction layer takes $\hat a$ as input and classify the image as $\hat y = f_c(\hat a) = f_c(f_a(v))$. We train the explainable layer $f_e(\cdot)$ and the final prediction $f_c(\cdot)$ jointly with the following loss:
\begin{equation}
\label{eq:dia loss}
\begin{aligned}
    L_{dia} &= \beta \cdot L_{e}(\hat a, a)+ L_{c}(\hat y, y) \\
    &= - \beta \cdot \frac{1}{N}\sum_{i=1}^{N}a_i \cdot log(\hat a_i) + (1-a_i) \cdot (1-log(\hat a_i)) - \frac{1}{M}\sum_{i=1}^{M}y_i \cdot log(\hat y_i)
\end{aligned} 
\end{equation}
Where $L_e(\cdot)$ denotes the loss of the explainable layer while $L_c(\cdot)$ the loss of the ED prognosis prediction. $\beta$ is a factor that balances between the two component losses. $a$ refers to the ground-truth causal explanation annotation for a given image and $\hat a$ is the prediction by $f_e(\cdot)$. $N$ represents the amount of the causal anomalies. Since it is a multi-label classification task, $L_e(\cdot)$ calculates the binary cross-entropy loss of the prediction and the ground-truth. Similarly, $y$ is the ground-truth label of ED prognosis and $\hat y$ is the predicted label using $f_c(\cdot)$. $L_c(\cdot)$ represents the cross-entropy loss between $y$ and $\hat y$. $M$ denotes the number of classes. 

We trained our model on the eye gaze enhanced chest X-ray image dataset (CXR-Eye) \cite{karargyris2021creation}. It contains 1083 chest X-ray images from the MIMIC-CXR dataset \cite{johnson2019mimic}. For each image, there is a causal explanation $a$ and each entry in $a$ represents an anomaly (disease), 1 denotes this disease is detected in this image while 0 not, and -1 means not certain. To simplify the question, -1 is also replaced by 0 \cite{johnson2019mimic}. There are in total 13 anomalies. These anomalies are extracted and labelled automatically using the tool CheXpert \cite{irvin2019chexpert}. $y$ is the class out of the three: pneumonia, congestive heart failure (CHF) and normal, which is the ED prognosis \cite{johnson2019mimiciv}.

Concretely, we used EfficientNet-b0 \cite{tan2019efficientnet} as the backbone of the encoder $\mathcal{E}(\cdot)$. $f_e(\cdot)$ contains one fully-connected (fc) and one activation layer, while $f_c(\cdot)$ contains four fc layers. We trained the model using an Adam optimizer for max. 100 epochs and the input data $X$ was resized to $\mathcal{R}^{3 \times 224 \times 224}$. To evaluate the model performance, we report the classification accuracy of the ED prognosis classifier and the Area Under Curve (AUC) of the causal explanations.
Since the dataset has a very limited number of data, we ran a 5-fold cross-validation on our proposed model and use the averaged score from five runs as the final score. We chose $\beta=1$, since it achieved the best performance with 74.12\% accuracy and 0.67 AUC score.

\subsection{User Study Design}
\begin{figure}[t]
  \centering
  \includegraphics[width=.7\linewidth]{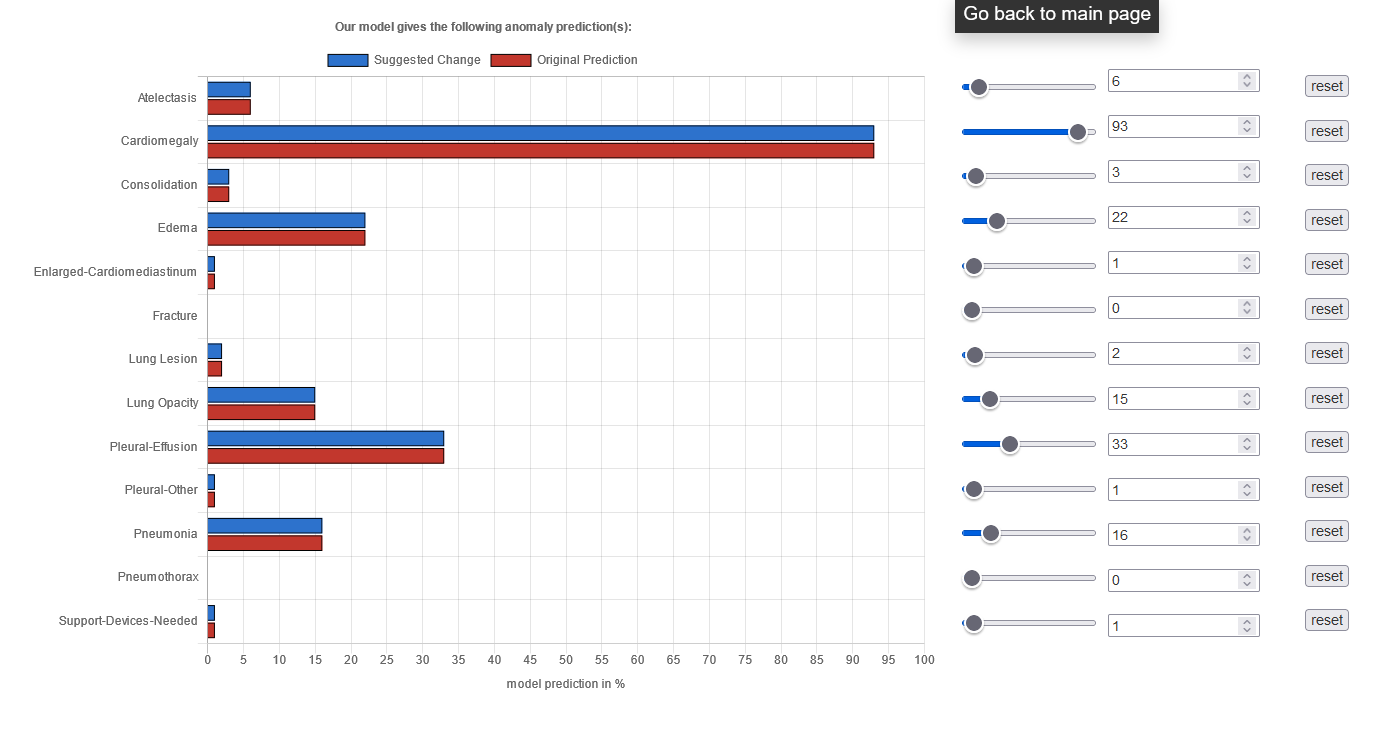}
  \caption{User interface for investigating the model's causal explanations. Users could alter the probability of the each causal explanation.}
  \label{fig:slide bar}
\end{figure}

\paragraph{Procedure.}
Our user study was designed in close collaboration with medical professionals from University Hospital Tübingen and conducted as an online questionnaire \footnote{The survey is at \url{http://hci-projects.informatik.uni-tuebingen.de/chiws}.}. 
Images and predictions used in the user study were from the test set of the model. Each participant was asked to answer the questions regarding the model performance: First, they chose whether they agreed with the model ED prognosis (pneumonia, CHF, or normal). Then, they viewed the causal explanations and could use a slider to change the probabilities of each of the 13 causal anomalies. These changes were shown as a red bar next to the model's original predictions in blue. There was an additional box for comments, where participants had the option to elaborate on their opinions (see \Cref{fig:slide bar}). 

After viewing all images, they rated scores using a 5-point Likert scale ranging from one ("strongly disagree") to five ("strongly agree") on five statements in \Cref{tab:trust factors} as self-reported trust measurement. We designed our statements according to the different factors of trust, some of which also give insights into the usability of the model such as personal attachment. 
Two experts from the medical domain helped us adapt the questions in order to find out experts' opinions of our model performance and function implementation.

\paragraph{Hypothesis.} 
We developed one main hypothesis on the expert trust in AI. In particular, we expect that users trust our model for their decision tasks on disease prediction and causal explanations. Our model was considerably accurate in diagnosis prediction on the test set with $74\%$ accuracy (chance level being $33\%$). We did not divulge this performance information to the radiologists, but we anticipated that they converge on similar decisions as with the model's predictions (i.e., appropriate trust level). 

\label{sec:study design}

\begin{table}[b]
\begin{tabular}{|c|c|c|}
\hline
 \textbf{Factor} & \textbf{Description} & \textbf{Question}  \\ \hline
 
 perceived understandability & \begin{tabular}[c]{@{}l@{}} user can understand the model \\ and predict its future behavior.
 \end{tabular} &  \begin{tabular}[c]{@{}l@{}} No need to learn a lot of things before \\ getting going with this AI model.  \end{tabular}\\ \hline
 perceived technical competence & \begin{tabular}[c]{@{}l@{}} model performs the tasks accurately and \\ correctly based on the input information.\end{tabular}  & \begin{tabular}[c]{@{}l@{}} This AI model \\ functioned well.\end{tabular} \\ \hline
 perceived reliability & \begin{tabular}[c]{@{}l@{}} model is in the sense of \\ repeated, consistent functioning.\end{tabular} & \begin{tabular}[c]{@{}l@{}} The AI model feedback of \\the Xrays was consistent. \end{tabular}\\ \hline

 personal attachment & \begin{tabular}[c]{@{}l@{}} user finds the model suits their taste \\ has a strong preference for it. \end{tabular} & \begin{tabular}[c]{@{}l@{}}The AI model was easy to use \\ and felt confident using it. \end{tabular} \\ \hline
 faith & \begin{tabular}[c]{@{}l@{}} user has faith in the future ability of the \\ model to perform even in unseen situations.\end{tabular}  & \begin{tabular}[c]{@{}l@{}}  I would like to use such a kind of \\ AI model in future work. \end{tabular}  \\ \hline
\end{tabular}
\caption{Main factors in trust \cite{madsen2000measuring} and corresponding questions in our user study.}
\label{tab:trust factors}
\end{table}

\paragraph{Measures and Analysis.} 
To measure the observed trust, we used the agreement rate (the percentage of times where users agreed with the model prediction) as the first indicator. As a second indicator, we analyzed each causal explanation given by users and model to see how much they differed from each other. To do this, we applied Wilcoxon signed-rank tests~\cite{wilcoxon1992individual}, which is a non-parametric equivalent of paired T-test, by using a significance level of $\alpha = 0.05$. The intuition behind the two indicators is that a higher agreement rate and similar distribution imply higher trust and vice versa. For each self-reported trust factor (listed in \Cref{tab:trust factors}), we calculated mean scores and their standard deviations.

A subsequent analysis was carried out to determine whether users were over-reliant or self-reliant on our model by dividing the data into two groups: (a) the model made a correct prediction and (b) the model made a wrong prediction. Inside each group, we ran Wilcoxon signed-rank tests on each causal explanation to see whether users were convinced by the explanations or not. Following the definition in \cite{bussone2015role}, if the alignment in (b) was high ($p < 0.05$), users were over-reliant; If the alignment in (a) was low, users were self-reliant. If the alignment in (b) was lower than (a), the users calibrated their trust fairly.

\paragraph{Participants.}
The web survey was given to the collaborating medical professionals who administered it to their co-workers. Five professionals working in the clinics participated. Two professionals had more than ten years of experience in radiology and the others had less than five years of working experience. None of them had experience working with AI models in their daily work routine. Each participant received a link having six images to investigate. This decision was to encourage experts to participate, even for just a short time, as it is hard for medical experts from clinics to set aside large time windows.

\begin{table*}[]
\resizebox{\textwidth}{!}{
\begin{tabular}{c|c|c|c|c|c|c|c|c|c|c|c|c|c}
\hline
 & Atelectasis & \begin{tabular}[c]{@{}c@{}}Cardio-  \\ megaly\end{tabular} & \begin{tabular}[c]{@{}c@{}}Consoli-  \\ dation\end{tabular} & Edema & \begin{tabular}[c]{@{}c@{}}Enlarged \\ Cardio.\end{tabular} & Fracture & \begin{tabular}[c]{@{}c@{}}Lung \\ Lesion\end{tabular} & \begin{tabular}[c]{@{}c@{}}Lung \\ Opacity\end{tabular} & \begin{tabular}[c]{@{}c@{}}Pleural\\ Effusion\end{tabular} & \begin{tabular}[c]{@{}c@{}}Pleural\\ Other\end{tabular} & Pneumonia & Pneumoth. & \begin{tabular}[c]{@{}c@{}}Support\\ Devices \end{tabular}\\ \hline
{Overall}   &   \begin{tabular}[c]{@{}c@{}}$p=0.150$ \\ $r=0.303$ \end{tabular} &   \begin{tabular}[c]{@{}c@{}}{$\bm{p=0.003}$} \\ $r=0.617$ \end{tabular} &  \begin{tabular}[c]{@{}c@{}}$p=0.284$ \\ $r=0.226$ \end{tabular}  & \begin{tabular}[c]{@{}c@{}}$\bm{p=0.038}$ \\ $r=0.437$ \end{tabular} & \begin{tabular}[c]{@{}c@{}}$p=0.789$ \\ $r=0.058$ \end{tabular}&  \begin{tabular}[c]{@{}c@{}}$p=0.365$ \\ $r=0.191$ \end{tabular} &  \begin{tabular}[c]{@{}c@{}}$p=0.821$ \\ $r=0.049$ \end{tabular} &  \begin{tabular}[c]{@{}c@{}}$\bm{p=0.0001}$ \\ $r=0.798$ \end{tabular} &  \begin{tabular}[c]{@{}c@{}}$\bm{p=0.003}$ \\ $r=0.613$ \end{tabular}  & \begin{tabular}[c]{@{}c@{}}$p=0.333$ \\ $r=0.204$ \end{tabular} & \begin{tabular}[c]{@{}c@{}}$p=0.323$ \\ $r=0.209$ \end{tabular} & \begin{tabular}[c]{@{}c@{}}$p=0.058$ \\ $r=0.398$ \end{tabular} &  \begin{tabular}[c]{@{}c@{}}$p=0.551$ \\ $r=0.127$ \end{tabular} \\ \hline\hline

 \begin{tabular}[c]{@{}c@{}}Correct\\ Prediction\end{tabular}   &      \begin{tabular}[c]{@{}c@{}}$p=0.320$ \\ $r=0.255$ \end{tabular} & \begin{tabular}[c]{@{}c@{}}{$\bm{p=0.001}$} \\ $r=0.792$ \end{tabular}  &    \begin{tabular}[c]{@{}c@{}}$\bm{p=0.046}$ \\ $r=0.498$ \end{tabular}  & \begin{tabular}[c]{@{}c@{}}$\bm{p=0.013}$ \\ $r=0.610$ \end{tabular} &   \begin{tabular}[c]{@{}c@{}}$p=0.203$ \\ $r=0.325$ \end{tabular} &      \begin{tabular}[c]{@{}c@{}}$p=0.055$ \\ $r=0.481$ \end{tabular}    & \begin{tabular}[c]{@{}c@{}}$p=0.759$ \\ $r=0.082$ \end{tabular}    & \begin{tabular}[c]{@{}c@{}}$\bm{p=0.013}$ \\ $r=0.610$ \end{tabular}  &   \begin{tabular}[c]{@{}c@{}}$\bm{p=0.004}$ \\ $r=0.697$ \end{tabular} &  \begin{tabular}[c]{@{}c@{}}$p=0.103$ \\ $r=0.411$ \end{tabular}  &  \begin{tabular}[c]{@{}c@{}}$p=0.147$ \\ $r=0.368$ \end{tabular}  & \begin{tabular}[c]{@{}c@{}}$p=0.055$ \\ $r=0.481$ \end{tabular}  & \begin{tabular}[c]{@{}c@{}}$p=0.759$ \\ $r=0.08$ \end{tabular}  \\ \hline
   \begin{tabular}[c]{@{}c@{}}Incorrect\\ Prediction\end{tabular} &  \begin{tabular}[c]{@{}c@{}}$p=0.426$ \\ $r=0.333$ \end{tabular}  & \begin{tabular}[c]{@{}c@{}}$p=0.910$ \\ $r=0.067$ \end{tabular} & \begin{tabular}[c]{@{}c@{}}$p=0.426$ \\ $r=0.333$ \end{tabular}  &  \begin{tabular}[c]{@{}c@{}}$p=0.570$ \\ $r=0.244$ \end{tabular} & \begin{tabular}[c]{@{}c@{}}$p=0.426$ \\ $r=0.333$ \end{tabular}  & \begin{tabular}[c]{@{}c@{}}$p=0.301$ \\ $r=0.422$ \end{tabular}  & \begin{tabular}[c]{@{}c@{}}$p=1.000$ \\ $r=0.022$ \end{tabular}   & \begin{tabular}[c]{@{}c@{}}$\bm{p=0.004}$ \\ $r=1.000$ \end{tabular} &\begin{tabular}[c]{@{}c@{}}$p=0.469$ \\ $r=0.289$ \end{tabular}  & \begin{tabular}[c]{@{}c@{}}$p=0.426$ \\ $r=0.333$ \end{tabular}   & \begin{tabular}[c]{@{}c@{}}$p=0.652$ \\ $r=0.200$ \end{tabular}  & \begin{tabular}[c]{@{}c@{}}$p=0.570$ \\ $r=0.244$ \end{tabular} & \begin{tabular}[c]{@{}c@{}}$p=0.652$ \\ $r=0.200$ \end{tabular}  \\ \hline
\end{tabular}
}
\caption{Results of the statistical analyses. For each causal explanation, we provide $p$ values using Wilcoxon signed-rank test~\cite{wilcoxon1992individual} and effect sizes with $r$ (rank biserial correlation~\cite{kerby2014simple}).}
\label{tab:wilcoxon}
\end{table*}

\section{Results}
In this section, we show the results based on the data we collected and analyze whether users trust our model considering our hypothesis. 

\subsection{Trust}
\paragraph{Observed Trust.} To test our hypothesis, we first calculated the agreement rate between the final predictions from our model and human participants. The agreement rate was $\mathbf{46.4\%}$. We then studied the second indicator and the results are listed in \Cref{tab:wilcoxon}. In the first row, we show the overall distribution similarity between causal explanations given by users and our model. In most cases $p > 0.05$, thus the null-hypothesis is rejected. For some of the anomalies, such as ``Cardiomegaly'', ``Lung Opacity'', and ``Pleural Effusion'', the $p$-values are either slightly below $0.05$ or in the vicinity of $0.05$ (e.g., for ``Edema'' $p~\approx 0.04$). 
Overall, users were observed to align with our model explanations. 

\begin{figure}[h]
     \centering
     \begin{subfigure}[]{0.4\textwidth}
         \centering
         \includegraphics[width=\textwidth]{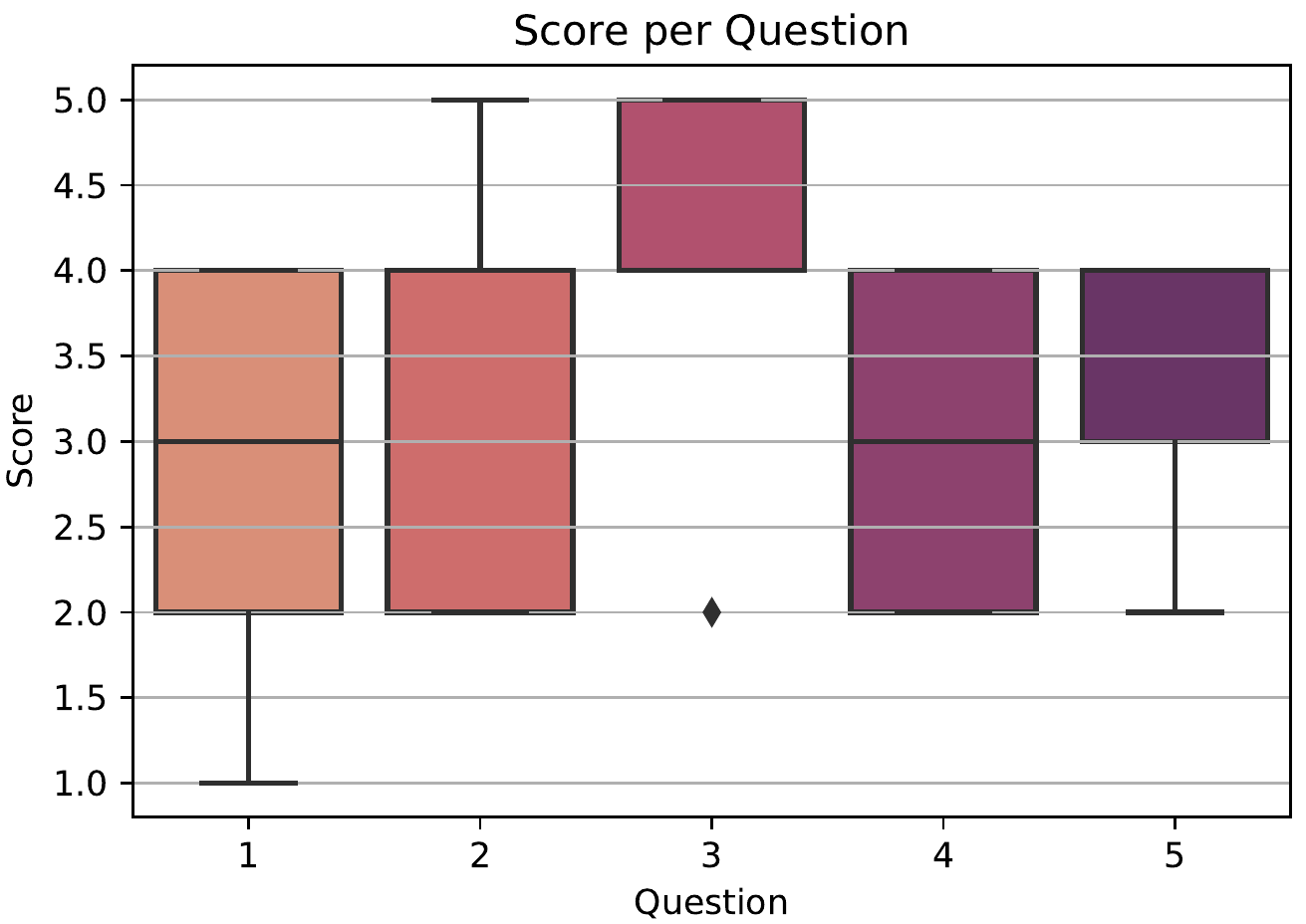}
         \caption{Scores for each question from all users.}
         \label{fig:score per question}
     \end{subfigure}
     \hspace{20.0pt}
     \begin{subfigure}[]{0.4\textwidth}
         \centering
         \includegraphics[width=\textwidth]{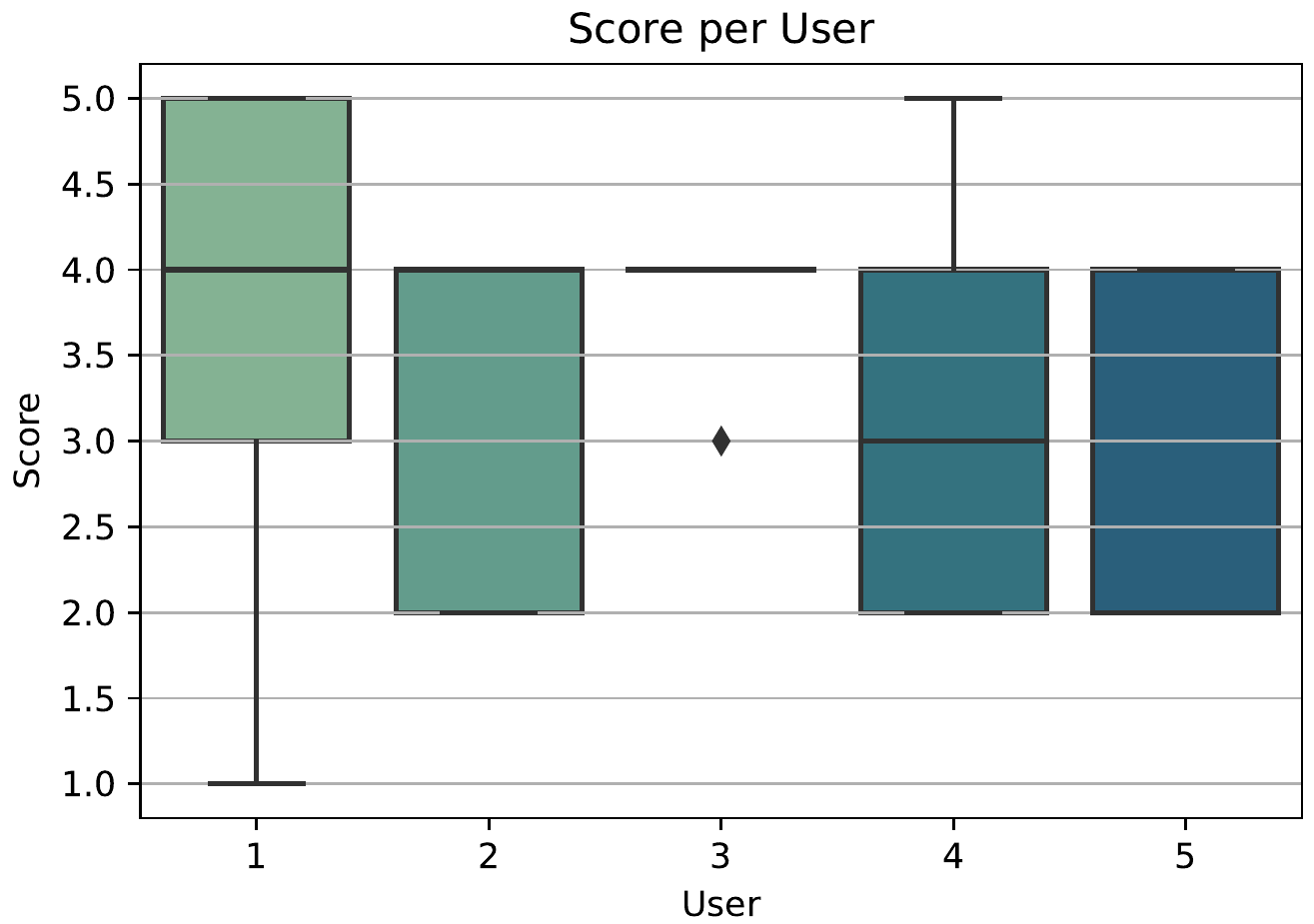}
         \caption{Scores for all questions from each user.}
         \label{fig:score per user}
     \end{subfigure}
        \caption{Self-reported scores for trust questions.}
        \label{fig:self-reported}
\end{figure}

\paragraph{Self-reported Trust.} In \Cref{tab:trust factors}, we proposed five questions to measure trust from different perspectives. The overall trust score (in the range of 1.0 to 5.0) is calculated as the median value among five averaged trust scores (following the evaluation in \cite{larasati2020effect}), which resulted in $3.2$. This score indicates that users overall trusted the model, but there is still the need to improve the trust. The scores in detail are presented in \Cref{fig:self-reported}. In \Cref{fig:score per question}, the scores for each question regarding one trust factor are illustrated. We see that the users gave positive feedback on the factors such as ``perceived reliability'', ``perceived understandability'' and ``faith in the future use'' ($Mdn = 4$). However, for the ``perceived technical competence'' and ``personal attachment'', users were not very confident ($Mdn = 3$). Different users seemed to have different opinions, which is depicted in \Cref{fig:score per user}. User 1, 2 and 3 rated every question with higher scores, while User 5 gave more negative feedback ($Mdn = 2$), indicating that this user was not convinced by the model. 

\subsection{Reliance}
To estimate which type of reliance users had, we studied user behavior when the model predicted correctly and when it did incorrectly. The results are shown in the bottom part of \Cref{tab:wilcoxon}. Note that users were not aware of whether the model was correct or not. In all the cases except for Lung Opacity, users aligned with the explanations from wrong predictions ($p > 0.05$), which indicates that users were over-reliant on the model explanations. Users trusted the explanations for correct predictions less often, for example in Cardiomegaly, Edema, Lung Opacity and Pleural Effusion, users tended to disagree ($p < 0.05$). Moreover, the users did not calibrate their trust fairly regarding Cardiomegaly, Edema and Pleural Effusion, since they trusted more in the explanations for wrong predictions than for correct predictions. 

\section{Discussion}


The radiologists' agreement to the model rate was only $\mathbf{46.4\%}$, indicating that users had very low trust in our model, even though it had good performance on the dataset. Our model performed with an accuracy of $\approx 75\%$ on the same images used in the user study. Users were not aware of this performance level and thus perceived it as lower performing. However, in the statistical analyses on the model explanations and the changes the radiologists made to them, we found that distributions differed only in few causal explanations. Furthermore, users tended to be over-reliant on some explanations. This result implies that users seemed to agree with most of the causal explanations provided by the model, while they were not convinced by the final prediction. These two paradoxical findings reveal that users had mixed feelings of trust and doubt in the model. The self-reported trust score, 3.2 out of 5, also validates that users leaned toward agreeing with the model predictions but not completely, suggesting they feel that the model performance should be further improved. Then, higher self-reported trust scores could be achieved, which requires further investigation both from machine learning and user-study perspective as only few misclassifications can deteriorate user trust heavily, especially in critical applications such as the medical diagnosis.

As none of the users had worked with AI models before, they were not able to calibrate their expectations or trust fairly. In this sense, model causal explanations did not have a perceivable effect on trust calibration, which is also in line with previous work~\cite{bussone2015role,zhang2020effect}. 
One possible reason for low trust could be the uncertainty in medical data, which leads to in general different opinions. For example, Schaekermann et al.~\cite{schaekermann2020ambiguity} support that when there is uncertainty in the medical it can affect experts' perception of its output. However, uncertainty is typical in medical diagnosis, even between radiologists~\cite{bruno2015understanding,kim2014fool,benchoufi2020interobserver}. How the uncertainty of the model needs to be conveyed is highly crucial to developing user trust. Generally, humans can pick up on the level of confidence in another human, but this intuition is not as clear from an AI~\cite{schaekermann2020ambiguity,amershi2019guidelines}. Our work conveys the model's probability that a certain anomaly will be diagnosed and this, coupled with the priming to help train the model, led the user to disagree more with the system. Therefore, better interaction between AI and experts should promote expert control and guidance, yet clearly convey where the model's level of ability. One way of solving this issue is to train users in the medical domain to work with AI models so that they can use these models as support in their daily routines for diagnostic purposes. It may be wise to encourage medical professionals to collaborate with such systems as they would with another colleague, determining a diagnosis from a second opinion rather than blinding acceptance. \cite{bussone2015role} expressed this similar opinion towards AIMDSS: ``Working in general practice is a hard job. I sit here on my own. I have to use my own knowledge. So this [system] is like having another person.'' Therefore, logical explanations are essential for users to understand models so that such AI models are not perceived as a black box.

However, there are several limitations that can be improved in future work. One limitation of our work is the number of experts contributing to the user study. Since medical experts are hard to acquire due to their extremely demanding schedules, a small number of experts is often the case in expertise research. One scoping review on expertise evaluated 73 sources and points out that the majority of these studies evaluate five, maybe ten experts\cite{gegenfurtner2011expertise}. 
Nevertheless, with more medical experts we could design a larger between-group user study to explore the effects of explanations in different qualities on user trust, which promotes interdisciplinary connections for better efforts towards integrating these models into professional environments.
Moreover, the dataset we used for training, CXR-Eye, is rather small. In order to have a better performing model, we plan to use knowledge distillation methods \cite{hinton2015distilling,mirzadeh2020improved} to transfer the knowledge from a well-performing model trained on a larger chest X-Ray dataset such as MIMIC-CXR \cite{johnson2019mimic}.

\section{Conclusion}
In this preliminary study, we designed an explainable AI model for X-Ray diagnosis, which provided causal explanations for the final prediction. We ran a user study to investigate whether the users trusted the model by collecting their perceived trust in the system as well as their agreement (observed trust) to both model prediction and explanations. In our analysis, we measured the observed trust using two indicators. The first one was the user agreement rate of the model prediction. Although our model achieved 74.12\% on the test set, users did not agree with the model final prediction in 54\% cases in the user study, suggesting that users did not trust the model enough. The second analysis was the Wilcoxon signed-ranked test on each causal explanation given by users and the model, where the result indicated that users were observed to align with our model explanations, even showing over-reliance on the explanations. Subsequently, perceived user trust was a self-report, where users rated the system as 3.2 out of 5, which validates that users were not confident in the model's ability.

\section{Acknowledgments}
We thank our medical experts from University Hospital Tübingen, Asklepios Clinic St. Georg, and Birkle-Clinic in Überlingen who participated in our user studies. We acknowledge support by the Deutsche Forschungsgemeinschaft (DFG, German Research Foundation), Cluster of Excellence - Machine Learning: New Perspectives for Science, EXC number 2064/1 - Project number 390727645. 
\bibliographystyle{ACM-Reference-Format}
\bibliography{sample-base}


\begin{thebibliography}{43}


\ifx \showCODEN    \undefined \def \showCODEN     #1{\unskip}     \fi
\ifx \showDOI      \undefined \def \showDOI       #1{#1}\fi
\ifx \showISBNx    \undefined \def \showISBNx     #1{\unskip}     \fi
\ifx \showISBNxiii \undefined \def \showISBNxiii  #1{\unskip}     \fi
\ifx \showISSN     \undefined \def \showISSN      #1{\unskip}     \fi
\ifx \showLCCN     \undefined \def \showLCCN      #1{\unskip}     \fi
\ifx \shownote     \undefined \def \shownote      #1{#1}          \fi
\ifx \showarticletitle \undefined \def \showarticletitle #1{#1}   \fi
\ifx \showURL      \undefined \def \showURL       {\relax}        \fi
\providecommand\bibfield[2]{#2}
\providecommand\bibinfo[2]{#2}
\providecommand\natexlab[1]{#1}
\providecommand\showeprint[2][]{arXiv:#2}

\bibitem[\protect\citeauthoryear{Akbarian, Seyyed-Kalantari, Khalvati, and
  Dolatabadi}{Akbarian et~al\mbox{.}}{2020}]%
        {akbarian2020evaluating}
\bibfield{author}{\bibinfo{person}{Sina Akbarian}, \bibinfo{person}{Laleh
  Seyyed-Kalantari}, \bibinfo{person}{Farzad Khalvati}, {and}
  \bibinfo{person}{Elham Dolatabadi}.} \bibinfo{year}{2020}\natexlab{}.
\newblock \showarticletitle{Evaluating knowledge transfer in neural network for
  medical images}.
\newblock \bibinfo{journal}{\emph{arXiv preprint arXiv:2008.13574}}
  (\bibinfo{year}{2020}).
\newblock


\bibitem[\protect\citeauthoryear{Amershi, Weld, Vorvoreanu, Fourney, Nushi,
  Collisson, Suh, Iqbal, Bennett, Inkpen, et~al\mbox{.}}{Amershi
  et~al\mbox{.}}{2019}]%
        {amershi2019guidelines}
\bibfield{author}{\bibinfo{person}{Saleema Amershi}, \bibinfo{person}{Dan
  Weld}, \bibinfo{person}{Mihaela Vorvoreanu}, \bibinfo{person}{Adam Fourney},
  \bibinfo{person}{Besmira Nushi}, \bibinfo{person}{Penny Collisson},
  \bibinfo{person}{Jina Suh}, \bibinfo{person}{Shamsi Iqbal},
  \bibinfo{person}{Paul~N Bennett}, \bibinfo{person}{Kori Inkpen},
  {et~al\mbox{.}}} \bibinfo{year}{2019}\natexlab{}.
\newblock \showarticletitle{Guidelines for human-AI interaction}. In
  \bibinfo{booktitle}{\emph{Proceedings of the 2019 chi conference on human
  factors in computing systems}}. \bibinfo{pages}{1--13}.
\newblock


\bibitem[\protect\citeauthoryear{Benchoufi, Matzner-Lober, Molinari, Jannot,
  and Soyer}{Benchoufi et~al\mbox{.}}{2020}]%
        {benchoufi2020interobserver}
\bibfield{author}{\bibinfo{person}{M Benchoufi}, \bibinfo{person}{E
  Matzner-Lober}, \bibinfo{person}{N Molinari}, \bibinfo{person}{A-S Jannot},
  {and} \bibinfo{person}{P Soyer}.} \bibinfo{year}{2020}\natexlab{}.
\newblock \showarticletitle{Interobserver agreement issues in radiology}.
\newblock \bibinfo{journal}{\emph{Diagnostic and Interventional Imaging}}
  \bibinfo{volume}{101}, \bibinfo{number}{10} (\bibinfo{year}{2020}),
  \bibinfo{pages}{639--641}.
\newblock


\bibitem[\protect\citeauthoryear{Bruno, Walker, and Abujudeh}{Bruno
  et~al\mbox{.}}{2015}]%
        {bruno2015understanding}
\bibfield{author}{\bibinfo{person}{Michael~A Bruno}, \bibinfo{person}{Eric~A
  Walker}, {and} \bibinfo{person}{Hani~H Abujudeh}.}
  \bibinfo{year}{2015}\natexlab{}.
\newblock \showarticletitle{Understanding and confronting our mistakes: the
  epidemiology of error in radiology and strategies for error reduction}.
\newblock \bibinfo{journal}{\emph{Radiographics}} \bibinfo{volume}{35},
  \bibinfo{number}{6} (\bibinfo{year}{2015}), \bibinfo{pages}{1668--1676}.
\newblock


\bibitem[\protect\citeauthoryear{Bu{\c{c}}inca, Malaya, and
  Gajos}{Bu{\c{c}}inca et~al\mbox{.}}{2021}]%
        {buccinca2021trust}
\bibfield{author}{\bibinfo{person}{Zana Bu{\c{c}}inca},
  \bibinfo{person}{Maja~Barbara Malaya}, {and} \bibinfo{person}{Krzysztof~Z
  Gajos}.} \bibinfo{year}{2021}\natexlab{}.
\newblock \showarticletitle{To trust or to think: cognitive forcing functions
  can reduce overreliance on AI in AI-assisted decision-making}.
\newblock \bibinfo{journal}{\emph{Proceedings of the ACM on Human-Computer
  Interaction}} \bibinfo{volume}{5}, \bibinfo{number}{CSCW1}
  (\bibinfo{year}{2021}), \bibinfo{pages}{1--21}.
\newblock


\bibitem[\protect\citeauthoryear{Budd, Robinson, and Kainz}{Budd
  et~al\mbox{.}}{2021}]%
        {budd2021survey}
\bibfield{author}{\bibinfo{person}{Samuel Budd}, \bibinfo{person}{Emma~C
  Robinson}, {and} \bibinfo{person}{Bernhard Kainz}.}
  \bibinfo{year}{2021}\natexlab{}.
\newblock \showarticletitle{A survey on active learning and human-in-the-loop
  deep learning for medical image analysis}.
\newblock \bibinfo{journal}{\emph{Medical Image Analysis}}
  (\bibinfo{year}{2021}), \bibinfo{pages}{102062}.
\newblock


\bibitem[\protect\citeauthoryear{Bussone, Stumpf, and O'Sullivan}{Bussone
  et~al\mbox{.}}{2015}]%
        {bussone2015role}
\bibfield{author}{\bibinfo{person}{Adrian Bussone}, \bibinfo{person}{Simone
  Stumpf}, {and} \bibinfo{person}{Dympna O'Sullivan}.}
  \bibinfo{year}{2015}\natexlab{}.
\newblock \showarticletitle{The role of explanations on trust and reliance in
  clinical decision support systems}. In \bibinfo{booktitle}{\emph{2015
  international conference on healthcare informatics}}. IEEE,
  \bibinfo{pages}{160--169}.
\newblock


\bibitem[\protect\citeauthoryear{Gegenfurtner, Lehtinen, and
  S{\"a}lj{\"o}}{Gegenfurtner et~al\mbox{.}}{2011}]%
        {gegenfurtner2011expertise}
\bibfield{author}{\bibinfo{person}{Andreas Gegenfurtner}, \bibinfo{person}{Erno
  Lehtinen}, {and} \bibinfo{person}{Roger S{\"a}lj{\"o}}.}
  \bibinfo{year}{2011}\natexlab{}.
\newblock \showarticletitle{Expertise differences in the comprehension of
  visualizations: A meta-analysis of eye-tracking research in professional
  domains}.
\newblock \bibinfo{journal}{\emph{Educational psychology review}}
  \bibinfo{volume}{23}, \bibinfo{number}{4} (\bibinfo{year}{2011}),
  \bibinfo{pages}{523--552}.
\newblock


\bibitem[\protect\citeauthoryear{Gunning and Aha}{Gunning and Aha}{2019}]%
        {gunning2019darpa}
\bibfield{author}{\bibinfo{person}{David Gunning} {and} \bibinfo{person}{David
  Aha}.} \bibinfo{year}{2019}\natexlab{}.
\newblock \showarticletitle{DARPA’s explainable artificial intelligence (XAI)
  program}.
\newblock \bibinfo{journal}{\emph{AI magazine}} \bibinfo{volume}{40},
  \bibinfo{number}{2} (\bibinfo{year}{2019}), \bibinfo{pages}{44--58}.
\newblock


\bibitem[\protect\citeauthoryear{Hinton, Vinyals, and Dean}{Hinton
  et~al\mbox{.}}{2015}]%
        {hinton2015distilling}
\bibfield{author}{\bibinfo{person}{Geoffrey Hinton}, \bibinfo{person}{Oriol
  Vinyals}, {and} \bibinfo{person}{Jeff Dean}.}
  \bibinfo{year}{2015}\natexlab{}.
\newblock \showarticletitle{Distilling the Knowledge in a Neural Network}.
\newblock \bibinfo{journal}{\emph{stat}}  \bibinfo{volume}{1050}
  (\bibinfo{year}{2015}), \bibinfo{pages}{9}.
\newblock


\bibitem[\protect\citeauthoryear{Holliday, Wilson, and Stumpf}{Holliday
  et~al\mbox{.}}{2016}]%
        {holliday2016user}
\bibfield{author}{\bibinfo{person}{Daniel Holliday}, \bibinfo{person}{Stephanie
  Wilson}, {and} \bibinfo{person}{Simone Stumpf}.}
  \bibinfo{year}{2016}\natexlab{}.
\newblock \showarticletitle{User trust in intelligent systems: A journey over
  time}. In \bibinfo{booktitle}{\emph{Proceedings of the 21st international
  conference on intelligent user interfaces}}. \bibinfo{pages}{164--168}.
\newblock


\bibitem[\protect\citeauthoryear{Irvin, Rajpurkar, Ko, Yu, Ciurea-Ilcus, Chute,
  Marklund, Haghgoo, Ball, Shpanskaya, et~al\mbox{.}}{Irvin
  et~al\mbox{.}}{2019}]%
        {irvin2019chexpert}
\bibfield{author}{\bibinfo{person}{Jeremy Irvin}, \bibinfo{person}{Pranav
  Rajpurkar}, \bibinfo{person}{Michael Ko}, \bibinfo{person}{Yifan Yu},
  \bibinfo{person}{Silviana Ciurea-Ilcus}, \bibinfo{person}{Chris Chute},
  \bibinfo{person}{Henrik Marklund}, \bibinfo{person}{Behzad Haghgoo},
  \bibinfo{person}{Robyn Ball}, \bibinfo{person}{Katie Shpanskaya},
  {et~al\mbox{.}}} \bibinfo{year}{2019}\natexlab{}.
\newblock \showarticletitle{Chexpert: A large chest radiograph dataset with
  uncertainty labels and expert comparison}. In
  \bibinfo{booktitle}{\emph{Proceedings of the AAAI conference on artificial
  intelligence}}, Vol.~\bibinfo{volume}{33}. \bibinfo{pages}{590--597}.
\newblock


\bibitem[\protect\citeauthoryear{Johnson, Lucas, Tom, Steven, Leo~Anthony, and
  Mark}{Johnson et~al\mbox{.}}{2021}]%
        {johnson2019mimiciv}
\bibfield{author}{\bibinfo{person}{Alistair Johnson},
  \bibinfo{person}{Bulgarelli Lucas}, \bibinfo{person}{Pollard Tom},
  \bibinfo{person}{Horng Steven}, \bibinfo{person}{Celi Leo~Anthony}, {and}
  \bibinfo{person}{Roger Mark}.} \bibinfo{year}{2021}\natexlab{}.
\newblock \showarticletitle{MIMIC-IV (version 1.0)}.
\newblock \bibinfo{journal}{\emph{PhysioNet}} (\bibinfo{year}{2021}).
\newblock


\bibitem[\protect\citeauthoryear{Johnson, Pollard, Berkowitz, Greenbaum,
  Lungren, Deng, Mark, and Horng}{Johnson et~al\mbox{.}}{2019}]%
        {johnson2019mimic}
\bibfield{author}{\bibinfo{person}{Alistair~EW Johnson}, \bibinfo{person}{Tom~J
  Pollard}, \bibinfo{person}{Seth~J Berkowitz}, \bibinfo{person}{Nathaniel~R
  Greenbaum}, \bibinfo{person}{Matthew~P Lungren}, \bibinfo{person}{Chih-ying
  Deng}, \bibinfo{person}{Roger~G Mark}, {and} \bibinfo{person}{Steven Horng}.}
  \bibinfo{year}{2019}\natexlab{}.
\newblock \showarticletitle{MIMIC-CXR, a de-identified publicly available
  database of chest radiographs with free-text reports}.
\newblock \bibinfo{journal}{\emph{Scientific data}} \bibinfo{volume}{6},
  \bibinfo{number}{1} (\bibinfo{year}{2019}), \bibinfo{pages}{1--8}.
\newblock


\bibitem[\protect\citeauthoryear{Karargyris, Kashyap, Lourentzou, Wu, Sharma,
  Tong, Abedin, Beymer, Mukherjee, Krupinski, et~al\mbox{.}}{Karargyris
  et~al\mbox{.}}{2021}]%
        {karargyris2021creation}
\bibfield{author}{\bibinfo{person}{Alexandros Karargyris},
  \bibinfo{person}{Satyananda Kashyap}, \bibinfo{person}{Ismini Lourentzou},
  \bibinfo{person}{Joy~T Wu}, \bibinfo{person}{Arjun Sharma},
  \bibinfo{person}{Matthew Tong}, \bibinfo{person}{Shafiq Abedin},
  \bibinfo{person}{David Beymer}, \bibinfo{person}{Vandana Mukherjee},
  \bibinfo{person}{Elizabeth~A Krupinski}, {et~al\mbox{.}}}
  \bibinfo{year}{2021}\natexlab{}.
\newblock \showarticletitle{Creation and validation of a chest X-ray dataset
  with eye-tracking and report dictation for AI development}.
\newblock \bibinfo{journal}{\emph{Scientific data}} \bibinfo{volume}{8},
  \bibinfo{number}{1} (\bibinfo{year}{2021}), \bibinfo{pages}{1--18}.
\newblock


\bibitem[\protect\citeauthoryear{Kerby}{Kerby}{2014}]%
        {kerby2014simple}
\bibfield{author}{\bibinfo{person}{Dave~S Kerby}.}
  \bibinfo{year}{2014}\natexlab{}.
\newblock \showarticletitle{The simple difference formula: An approach to
  teaching nonparametric correlation}.
\newblock \bibinfo{journal}{\emph{Comprehensive Psychology}}
  \bibinfo{volume}{3} (\bibinfo{year}{2014}), \bibinfo{pages}{11--IT}.
\newblock


\bibitem[\protect\citeauthoryear{Kim}{Kim}{2015}]%
        {kim2015interactive}
\bibfield{author}{\bibinfo{person}{Been Kim}.} \bibinfo{year}{2015}\natexlab{}.
\newblock \emph{\bibinfo{title}{Interactive and interpretable machine learning
  models for human machine collaboration}}.
\newblock \bibinfo{thesistype}{Ph.D. Dissertation}.
  \bibinfo{school}{Massachusetts Institute of Technology}.
\newblock


\bibitem[\protect\citeauthoryear{Kim and Mansfield}{Kim and Mansfield}{2014}]%
        {kim2014fool}
\bibfield{author}{\bibinfo{person}{Young~W Kim} {and} \bibinfo{person}{Liem~T
  Mansfield}.} \bibinfo{year}{2014}\natexlab{}.
\newblock \showarticletitle{Fool me twice: delayed diagnoses in radiology with
  emphasis on perpetuated errors}.
\newblock \bibinfo{journal}{\emph{American Journal of Roentgenology}}
  \bibinfo{volume}{202}, \bibinfo{number}{3} (\bibinfo{year}{2014}),
  \bibinfo{pages}{465--470}.
\newblock


\bibitem[\protect\citeauthoryear{Koh, Nguyen, Tang, Mussmann, Pierson, Kim, and
  Liang}{Koh et~al\mbox{.}}{2020}]%
        {koh2020concept}
\bibfield{author}{\bibinfo{person}{Pang~Wei Koh}, \bibinfo{person}{Thao
  Nguyen}, \bibinfo{person}{Yew~Siang Tang}, \bibinfo{person}{Stephen
  Mussmann}, \bibinfo{person}{Emma Pierson}, \bibinfo{person}{Been Kim}, {and}
  \bibinfo{person}{Percy Liang}.} \bibinfo{year}{2020}\natexlab{}.
\newblock \showarticletitle{Concept bottleneck models}. In
  \bibinfo{booktitle}{\emph{International Conference on Machine Learning}}.
  PMLR, \bibinfo{pages}{5338--5348}.
\newblock


\bibitem[\protect\citeauthoryear{Larasati, De~Liddo, and Motta}{Larasati
  et~al\mbox{.}}{2020}]%
        {larasati2020effect}
\bibfield{author}{\bibinfo{person}{Retno Larasati}, \bibinfo{person}{Anna
  De~Liddo}, {and} \bibinfo{person}{Enrico Motta}.}
  \bibinfo{year}{2020}\natexlab{}.
\newblock \showarticletitle{The Effect of Explanation Styles on User's Trust.}.
  In \bibinfo{booktitle}{\emph{ExSS-ATEC@ IUI}}.
\newblock


\bibitem[\protect\citeauthoryear{Lee and See}{Lee and See}{2004}]%
        {lee2004trust}
\bibfield{author}{\bibinfo{person}{John~D Lee} {and} \bibinfo{person}{Katrina~A
  See}.} \bibinfo{year}{2004}\natexlab{}.
\newblock \showarticletitle{Trust in automation: Designing for appropriate
  reliance}.
\newblock \bibinfo{journal}{\emph{Human factors}} \bibinfo{volume}{46},
  \bibinfo{number}{1} (\bibinfo{year}{2004}), \bibinfo{pages}{50--80}.
\newblock


\bibitem[\protect\citeauthoryear{Lipton}{Lipton}{2018}]%
        {lipton2018mythos}
\bibfield{author}{\bibinfo{person}{Zachary~C Lipton}.}
  \bibinfo{year}{2018}\natexlab{}.
\newblock \showarticletitle{The Mythos of Model Interpretability: In machine
  learning, the concept of interpretability is both important and slippery.}
\newblock \bibinfo{journal}{\emph{Queue}} \bibinfo{volume}{16},
  \bibinfo{number}{3} (\bibinfo{year}{2018}), \bibinfo{pages}{31--57}.
\newblock


\bibitem[\protect\citeauthoryear{Madsen and Gregor}{Madsen and Gregor}{2000}]%
        {madsen2000measuring}
\bibfield{author}{\bibinfo{person}{Maria Madsen} {and} \bibinfo{person}{Shirley
  Gregor}.} \bibinfo{year}{2000}\natexlab{}.
\newblock \showarticletitle{Measuring human-computer trust}. In
  \bibinfo{booktitle}{\emph{11th australasian conference on information
  systems}}, Vol.~\bibinfo{volume}{53}. Citeseer, \bibinfo{pages}{6--8}.
\newblock


\bibitem[\protect\citeauthoryear{Mirzadeh, Farajtabar, Li, Levine, Matsukawa,
  and Ghasemzadeh}{Mirzadeh et~al\mbox{.}}{2020}]%
        {mirzadeh2020improved}
\bibfield{author}{\bibinfo{person}{Seyed~Iman Mirzadeh},
  \bibinfo{person}{Mehrdad Farajtabar}, \bibinfo{person}{Ang Li},
  \bibinfo{person}{Nir Levine}, \bibinfo{person}{Akihiro Matsukawa}, {and}
  \bibinfo{person}{Hassan Ghasemzadeh}.} \bibinfo{year}{2020}\natexlab{}.
\newblock \showarticletitle{Improved knowledge distillation via teacher
  assistant}. In \bibinfo{booktitle}{\emph{Proceedings of the AAAI Conference
  on Artificial Intelligence}}, Vol.~\bibinfo{volume}{34}.
  \bibinfo{pages}{5191--5198}.
\newblock


\bibitem[\protect\citeauthoryear{Mohseni, Zarei, and Ragan}{Mohseni
  et~al\mbox{.}}{2021}]%
        {mohseni2021multidisciplinary}
\bibfield{author}{\bibinfo{person}{Sina Mohseni}, \bibinfo{person}{Niloofar
  Zarei}, {and} \bibinfo{person}{Eric~D Ragan}.}
  \bibinfo{year}{2021}\natexlab{}.
\newblock \showarticletitle{A multidisciplinary survey and framework for design
  and evaluation of explainable AI systems}.
\newblock \bibinfo{journal}{\emph{ACM Transactions on Interactive Intelligent
  Systems (TiiS)}} \bibinfo{volume}{11}, \bibinfo{number}{3-4}
  (\bibinfo{year}{2021}), \bibinfo{pages}{1--45}.
\newblock


\bibitem[\protect\citeauthoryear{Papenmeier, Englebienne, and
  Seifert}{Papenmeier et~al\mbox{.}}{2019}]%
        {papenmeier2019model}
\bibfield{author}{\bibinfo{person}{Andrea Papenmeier}, \bibinfo{person}{Gwenn
  Englebienne}, {and} \bibinfo{person}{Christin Seifert}.}
  \bibinfo{year}{2019}\natexlab{}.
\newblock \showarticletitle{How model accuracy and explanation fidelity
  influence user trust}.
\newblock \bibinfo{journal}{\emph{arXiv preprint arXiv:1907.12652}}
  (\bibinfo{year}{2019}).
\newblock


\bibitem[\protect\citeauthoryear{Pu and Chen}{Pu and Chen}{2006}]%
        {pu2006trust}
\bibfield{author}{\bibinfo{person}{Pearl Pu} {and} \bibinfo{person}{Li Chen}.}
  \bibinfo{year}{2006}\natexlab{}.
\newblock \showarticletitle{Trust building with explanation interfaces}. In
  \bibinfo{booktitle}{\emph{Proceedings of the 11th international conference on
  Intelligent user interfaces}}. \bibinfo{pages}{93--100}.
\newblock


\bibitem[\protect\citeauthoryear{Radensky, Downey, Lo, Popovi{\'c}, and
  Weld}{Radensky et~al\mbox{.}}{2021}]%
        {radensky2021exploring}
\bibfield{author}{\bibinfo{person}{Marissa Radensky}, \bibinfo{person}{Doug
  Downey}, \bibinfo{person}{Kyle Lo}, \bibinfo{person}{Zoran Popovi{\'c}},
  {and} \bibinfo{person}{Daniel~S Weld}.} \bibinfo{year}{2021}\natexlab{}.
\newblock \showarticletitle{Exploring The Role of Local and Global Explanations
  in Recommender Systems}.
\newblock \bibinfo{journal}{\emph{arXiv preprint arXiv:2109.13301}}
  (\bibinfo{year}{2021}).
\newblock


\bibitem[\protect\citeauthoryear{Rajpurkar, Irvin, Ball, Zhu, Yang, Mehta,
  Duan, Ding, Bagul, Langlotz, et~al\mbox{.}}{Rajpurkar et~al\mbox{.}}{2018}]%
        {rajpurkar2018deep}
\bibfield{author}{\bibinfo{person}{Pranav Rajpurkar}, \bibinfo{person}{Jeremy
  Irvin}, \bibinfo{person}{Robyn~L Ball}, \bibinfo{person}{Kaylie Zhu},
  \bibinfo{person}{Brandon Yang}, \bibinfo{person}{Hershel Mehta},
  \bibinfo{person}{Tony Duan}, \bibinfo{person}{Daisy Ding},
  \bibinfo{person}{Aarti Bagul}, \bibinfo{person}{Curtis~P Langlotz},
  {et~al\mbox{.}}} \bibinfo{year}{2018}\natexlab{}.
\newblock \showarticletitle{Deep learning for chest radiograph diagnosis: A
  retrospective comparison of the CheXNeXt algorithm to practicing
  radiologists}.
\newblock \bibinfo{journal}{\emph{PLoS medicine}} \bibinfo{volume}{15},
  \bibinfo{number}{11} (\bibinfo{year}{2018}), \bibinfo{pages}{e1002686}.
\newblock


\bibitem[\protect\citeauthoryear{Ribeiro, Singh, and Guestrin}{Ribeiro
  et~al\mbox{.}}{2016}]%
        {ribeiro2016should}
\bibfield{author}{\bibinfo{person}{Marco~Tulio Ribeiro},
  \bibinfo{person}{Sameer Singh}, {and} \bibinfo{person}{Carlos Guestrin}.}
  \bibinfo{year}{2016}\natexlab{}.
\newblock \showarticletitle{" Why should i trust you?" Explaining the
  predictions of any classifier}. In \bibinfo{booktitle}{\emph{Proceedings of
  the 22nd ACM SIGKDD international conference on knowledge discovery and data
  mining}}. \bibinfo{pages}{1135--1144}.
\newblock


\bibitem[\protect\citeauthoryear{Ribeiro, Singh, and Guestrin}{Ribeiro
  et~al\mbox{.}}{2018}]%
        {ribeiro2018anchors}
\bibfield{author}{\bibinfo{person}{Marco~Tulio Ribeiro},
  \bibinfo{person}{Sameer Singh}, {and} \bibinfo{person}{Carlos Guestrin}.}
  \bibinfo{year}{2018}\natexlab{}.
\newblock \showarticletitle{Anchors: High-precision model-agnostic
  explanations}. In \bibinfo{booktitle}{\emph{Proceedings of the AAAI
  conference on artificial intelligence}}, Vol.~\bibinfo{volume}{32}.
\newblock


\bibitem[\protect\citeauthoryear{Schaekermann, Beaton, Sanoubari, Lim, Larson,
  and Law}{Schaekermann et~al\mbox{.}}{2020}]%
        {schaekermann2020ambiguity}
\bibfield{author}{\bibinfo{person}{Mike Schaekermann}, \bibinfo{person}{Graeme
  Beaton}, \bibinfo{person}{Elaheh Sanoubari}, \bibinfo{person}{Andrew Lim},
  \bibinfo{person}{Kate Larson}, {and} \bibinfo{person}{Edith Law}.}
  \bibinfo{year}{2020}\natexlab{}.
\newblock \showarticletitle{Ambiguity-aware ai assistants for medical data
  analysis}. In \bibinfo{booktitle}{\emph{Proceedings of the 2020 CHI
  conference on human factors in computing systems}}. \bibinfo{pages}{1--14}.
\newblock


\bibitem[\protect\citeauthoryear{Schoonderwoerd, Jorritsma, Neerincx, and
  van~den Bosch}{Schoonderwoerd et~al\mbox{.}}{2021}]%
        {schoonderwoerd2021human}
\bibfield{author}{\bibinfo{person}{Tjeerd~AJ Schoonderwoerd},
  \bibinfo{person}{Wiard Jorritsma}, \bibinfo{person}{Mark~A Neerincx}, {and}
  \bibinfo{person}{Karel van~den Bosch}.} \bibinfo{year}{2021}\natexlab{}.
\newblock \showarticletitle{Human-Centered XAI: Developing Design Patterns for
  Explanations of Clinical Decision Support Systems}.
\newblock \bibinfo{journal}{\emph{International Journal of Human-Computer
  Studies}} (\bibinfo{year}{2021}), \bibinfo{pages}{102684}.
\newblock


\bibitem[\protect\citeauthoryear{Seyyed-Kalantari, Liu, McDermott, Chen, and
  Ghassemi}{Seyyed-Kalantari et~al\mbox{.}}{2020}]%
        {seyyed2020chexclusion}
\bibfield{author}{\bibinfo{person}{Laleh Seyyed-Kalantari},
  \bibinfo{person}{Guanxiong Liu}, \bibinfo{person}{Matthew McDermott},
  \bibinfo{person}{Irene~Y Chen}, {and} \bibinfo{person}{Marzyeh Ghassemi}.}
  \bibinfo{year}{2020}\natexlab{}.
\newblock \showarticletitle{CheXclusion: Fairness gaps in deep chest X-ray
  classifiers}. In \bibinfo{booktitle}{\emph{BIOCOMPUTING 2021: Proceedings of
  the Pacific Symposium}}. World Scientific, \bibinfo{pages}{232--243}.
\newblock


\bibitem[\protect\citeauthoryear{S{\"o}llner, Hoffmann, Hoffmann, and
  Leimeister}{S{\"o}llner et~al\mbox{.}}{2012}]%
        {sollner2012use}
\bibfield{author}{\bibinfo{person}{Matthias S{\"o}llner}, \bibinfo{person}{Axel
  Hoffmann}, \bibinfo{person}{Holger Hoffmann}, {and}
  \bibinfo{person}{Jan~Marco Leimeister}.} \bibinfo{year}{2012}\natexlab{}.
\newblock \showarticletitle{How to use behavioral research insights on trust
  for HCI system design}.
\newblock In \bibinfo{booktitle}{\emph{CHI'12 Extended Abstracts on Human
  Factors in Computing Systems}}. \bibinfo{pages}{1703--1708}.
\newblock


\bibitem[\protect\citeauthoryear{Tan and Le}{Tan and Le}{2019}]%
        {tan2019efficientnet}
\bibfield{author}{\bibinfo{person}{Mingxing Tan} {and} \bibinfo{person}{Quoc
  Le}.} \bibinfo{year}{2019}\natexlab{}.
\newblock \showarticletitle{Efficientnet: Rethinking model scaling for
  convolutional neural networks}. In \bibinfo{booktitle}{\emph{International
  Conference on Machine Learning}}. PMLR, \bibinfo{pages}{6105--6114}.
\newblock


\bibitem[\protect\citeauthoryear{Wang, Peng, Lu, Lu, Bagheri, and Summers}{Wang
  et~al\mbox{.}}{2017}]%
        {wang2017chestx}
\bibfield{author}{\bibinfo{person}{Xiaosong Wang}, \bibinfo{person}{Yifan
  Peng}, \bibinfo{person}{Le Lu}, \bibinfo{person}{Zhiyong Lu},
  \bibinfo{person}{Mohammadhadi Bagheri}, {and} \bibinfo{person}{Ronald~M
  Summers}.} \bibinfo{year}{2017}\natexlab{}.
\newblock \showarticletitle{Chestx-ray8: Hospital-scale chest x-ray database
  and benchmarks on weakly-supervised classification and localization of common
  thorax diseases}. In \bibinfo{booktitle}{\emph{Proceedings of the IEEE
  conference on computer vision and pattern recognition}}.
  \bibinfo{pages}{2097--2106}.
\newblock


\bibitem[\protect\citeauthoryear{Wilcoxon}{Wilcoxon}{1992}]%
        {wilcoxon1992individual}
\bibfield{author}{\bibinfo{person}{Frank Wilcoxon}.}
  \bibinfo{year}{1992}\natexlab{}.
\newblock \showarticletitle{Individual comparisons by ranking methods}.
\newblock In \bibinfo{booktitle}{\emph{Breakthroughs in statistics}}.
  \bibinfo{publisher}{Springer}, \bibinfo{pages}{196--202}.
\newblock


\bibitem[\protect\citeauthoryear{Xie, Chen, Kao, Gao, and Chen}{Xie
  et~al\mbox{.}}{2020}]%
        {xie2020chexplain}
\bibfield{author}{\bibinfo{person}{Yao Xie}, \bibinfo{person}{Melody Chen},
  \bibinfo{person}{David Kao}, \bibinfo{person}{Ge Gao}, {and}
  \bibinfo{person}{Xiang'Anthony' Chen}.} \bibinfo{year}{2020}\natexlab{}.
\newblock \showarticletitle{CheXplain: Enabling Physicians to Explore and
  Understand Data-Driven, AI-Enabled Medical Imaging Analysis}. In
  \bibinfo{booktitle}{\emph{Proceedings of the 2020 CHI Conference on Human
  Factors in Computing Systems}}. \bibinfo{pages}{1--13}.
\newblock


\bibitem[\protect\citeauthoryear{Yao, Poblenz, Dagunts, Covington, Bernard, and
  Lyman}{Yao et~al\mbox{.}}{2017}]%
        {yao2017learning}
\bibfield{author}{\bibinfo{person}{Li Yao}, \bibinfo{person}{Eric Poblenz},
  \bibinfo{person}{Dmitry Dagunts}, \bibinfo{person}{Ben Covington},
  \bibinfo{person}{Devon Bernard}, {and} \bibinfo{person}{Kevin Lyman}.}
  \bibinfo{year}{2017}\natexlab{}.
\newblock \showarticletitle{Learning to diagnose from scratch by exploiting
  dependencies among labels}.
\newblock \bibinfo{journal}{\emph{arXiv preprint arXiv:1710.10501}}
  (\bibinfo{year}{2017}).
\newblock


\bibitem[\protect\citeauthoryear{Yin, Wortman~Vaughan, and Wallach}{Yin
  et~al\mbox{.}}{2019}]%
        {yin2019understanding}
\bibfield{author}{\bibinfo{person}{Ming Yin}, \bibinfo{person}{Jennifer
  Wortman~Vaughan}, {and} \bibinfo{person}{Hanna Wallach}.}
  \bibinfo{year}{2019}\natexlab{}.
\newblock \showarticletitle{Understanding the effect of accuracy on trust in
  machine learning models}. In \bibinfo{booktitle}{\emph{Proceedings of the
  2019 chi conference on human factors in computing systems}}.
  \bibinfo{pages}{1--12}.
\newblock


\bibitem[\protect\citeauthoryear{Zhang, Li, Li, and Li}{Zhang
  et~al\mbox{.}}{2018}]%
        {zhang2018predicting}
\bibfield{author}{\bibinfo{person}{Wei Zhang}, \bibinfo{person}{Jun Li},
  \bibinfo{person}{Zu-Bing Li}, {and} \bibinfo{person}{Zhi Li}.}
  \bibinfo{year}{2018}\natexlab{}.
\newblock \showarticletitle{Predicting postoperative facial swelling following
  impacted mandibular third molars extraction by using artificial neural
  networks evaluation}.
\newblock \bibinfo{journal}{\emph{Scientific reports}} \bibinfo{volume}{8},
  \bibinfo{number}{1} (\bibinfo{year}{2018}), \bibinfo{pages}{1--9}.
\newblock


\bibitem[\protect\citeauthoryear{Zhang, Liao, and Bellamy}{Zhang
  et~al\mbox{.}}{2020}]%
        {zhang2020effect}
\bibfield{author}{\bibinfo{person}{Yunfeng Zhang}, \bibinfo{person}{Q~Vera
  Liao}, {and} \bibinfo{person}{Rachel~KE Bellamy}.}
  \bibinfo{year}{2020}\natexlab{}.
\newblock \showarticletitle{Effect of confidence and explanation on accuracy
  and trust calibration in AI-assisted decision making}. In
  \bibinfo{booktitle}{\emph{Proceedings of the 2020 Conference on Fairness,
  Accountability, and Transparency}}. \bibinfo{pages}{295--305}.
\newblock


\end{thebibliography}










\end{document}